# Photons Do Collapse In the Retina Not in the Brain Cortex: Evidence from Visual Illusions

### Danko Georgiev

**Abstract**

While looking for evidence of quantum coherent states within the brain, many quantum mind advocates proposed experiments based on the assumption that the coherent state of a photon entering the visual system could somehow be preserved through the neural processing, or in other words they suppose that photons collapse not in the retina, but in the brain cortex. In this paper we show that photons do collapse within the retina and subsequent processing of information at the level of neural membranes proceeds. Moreover, we explicitly stress on the fact that due to existent amplification of the signal produced by each photon, a basic quantum mechanical theorem forbids the photon state to be teleported from the retina to the brain cortex. The changes of the membrane potential of the neurons in the primary visual cortex are shown to be relevant to inputting visual sensory information that is already processed and is not identical to the visual image entering in the retina. A striking evidence for the existent processing of the incoming visual information by the retina is provided by visual illusions resulting from the lateral inhibition mechanism.

*"If you can't say it clearly, you don't understand it yourself."*

**John Searle**

## 1. Introduction

In the last two decades increasing number of researchers speculates that *consciousness*, which is often referred to as the last great mystery of science, should have *quantum* origins. However the amount of literature accumulating on the subject is either *bad science* or *pseudoscience*. The word "quantum" is used inappropriately by authors with little understanding of its actual meaning who are mostly interested in making their discourse sound more technical and scientific than it otherwise would be. Moreover, the very same writers have little or no knowledge of neuroscience and as a consequence their "quantum mind" proposals are already wrong at the moment of their conception. In this article we will introduce the reader to the basic neural principles underlying the *visual perception*, and then we will show that three putative quantum versions of visual perception proposed by (1) Hameroff's group (Hameroff, 1998; Saint Hilaire *et al*., 2002), (2) Gao (Gao, 2003; 2004a; 2004b; 2008) and (3) Tuszynski's group (Salari *et al*., 2008; Rahnama *et al*., 2009) are *not even wrong*[2]. The mistakes in the discussed works are numerous, so we will not claim to have been exhaustive in our critique. Instead we will pick up a few landmark features of each proposal, discuss what is wrong with them

Corresponding author: Danko Georgiev
Address: Graduate School of Medical Science, Kanazawa University, Kanazawa, Ishikawa 920-8641, Japan
e-mail: dankomed@yahoo.com

[2] The latter phrase was coined by the quantum physicist Wolfgang Pauli and suggests that even a wrong argument would have been better.

and then we provide guidelines for possible further scientific exploration. In other words, we will try to establish a viable quantum mind scientific research program in Lakatos sense[3] with a hope that future publications on the subject will be worth thorough reading and will not violate from the beginning the accumulated mountain of knowledge concerning the functioning of the visual system.

One of the most popular quantum mind models is the Orchestrated Objective Reduction (Orch OR) created by Hameroff and Penrose (1996). Striding towards achieving empirical accessibility, Hameroff (1998) provided a list of twenty testable predictions related to each of several critical statements of the Orch OR theory. One of the statements is that microtubule-based cilia/centriole structures are quantum optical devices: *"microtubule-based cilia in rods and cones directly detect visual photons and connect with retinal glial cell microtubule via gap junctions"*. Though it is eccentric to suppose that microtubule based cilia can capture visual photons and microtubules can act as waveguides to transmit the photons to the cerebral cortex, it is beyond comprehension why the retinal glia microtubules need to be involved in the process. Details of the proposal by Saint Hilaire *et al.* (2002) how to detect quantum coherent states in the retina via photon echo experiments are provided in *Section 3.1*.

Second putatively quantum mechanical theory advocated by Gao (2003; 2004a; 2004b; 2008) aims at measuring the quantum states of a qubit with the use of conscious observer. In particular it is proposed that sending coherently superposed photons into the eye of conscious observer is better than sending the superposed photons into a measuring apparatus. Gao claims that the consciousness might distinguish non-orthogonal states of the photons, while a measuring apparatus cannot; hence there is chance for superluminal communication if we use the unexplored power of consciousness. The fallacy in the latter proposition is discussed in *Section 3.2*.

And third, newly born model of visual perception involving quantum teleportation mechanism is proposed by Tuszynski's group (cf. Salari *et al.*, 2008; Rahnama *et al.*, 2009). The authors propose that the visual photons are quantum teleported to the brain cortex and they collapse there instead inside the retina. The flaws in such a quantum teleportation scheme are discussed in *Section 3.3*.

Before we address the possible quantum mechanisms in visual perception, we will introduce the reader to the current knowledge we already have from molecular neuroscience and neurophysiology. The basic structural organization of the visual pathways, the way the information is encoded in the form of electric signal, the possibility of visual illusions, as well as the various genetic abnormalities such as *color blindness*, all severely constraint the possible theories of visual perception. That is why we start from what we know about vision and proceed to show why the suggested three quantum proposals for visual perception cannot be true. At the end, we conclude the article with some general remarks on the quantum consciousness hypothesis, and explain why the current work should be viewed as a significant step forward in developing new quantum mind theories consistent with biological data.

## 2. Neurobiological basis of visual perception
### 2.1 Neurobiology of retinal transduction
Our eyes see the incoming light photons due to the existent *photoreceptors* located in the *retina*. The photoreceptors are responsible for the *transduction*[4] of light into electrical signals, which are then delivered to the brain

---

[3] A progressive research program includes a sequence of theories (Theory 1, Theory 2, ..., etc.), each one being consistent with all the known facts and predicting also new facts. In contrast, a degenerating research program is neither consistent with the known facts, nor the predicted by it new facts are confirmed. In the current essay we propose the "quantum mind" to be formulated as an ongoing research program in which new wider theories are to be further developed, though frankly speaking at present the quantum mind speculations are in a degenerating phase. According to Imre Lakatos *"One may rationally stick to a degenerating research programme until it is overtaken by a rival and even after. What one must not do is to deny its poor public record.... It is perfectly rational to play a risky game: what is irrational is to deceive oneself about the risk"* (Lakatos, 1971, p. 104).

[4] In physiology, the term *transduction* denotes the conversion of a stimulus from one form to another.

cortex, where we consciously perceive the visual images.

### 2.1.1 Photon transduction and signal amplification

The sensory transduction (the conversion of the incoming light into electric signal) takes place in the *photoreceptors* (rods and cones). It is a three stage process involving (1) photon induced isomerization of a pigment called *retinal*, subsequent to the absorption of a photon, (2) a biochemical cascade to amplify the incoming signal, and subsequent (3) alteration in the conductance of plasmalemmal[5] cyclic nucleotide- gated (CNG) ion channels permeable for $Na^+$ ions (see *Figure 1*).

Alterations in the ion currents lead to decrease or increase of the transmitter release (in this case glutamate) and thus affect the transfer of information to other neurons. The stacked disks in the *outer segment*[6] of rods and cones contain membrane proteins called *opsins*, which are members of the family of G-protein coupled receptors. Rods contain *rhodopsin* and are responsible for night vision. They have higher sensitivity for light compared to cone cells and are evolutionary younger (cf. Masland 2001). Theoretically it has been calculated that a rod in starlight could be activated even by absorption of a single photon. Cones are responsible for color vision and have higher requirements for light, which means that they are less sensitive in comparison to rods. Cones contain one of three types of photopsins: opsin R (red), opsin G (green), or opsin B (blue)[7], therefore there are red, green and blue cones[8]. Collectively the rhodopsin and the three types of photopsins are known as *photopigments*.

The larger fragment of each photopigment is a protein called *opsin*, which is a 7 transmembrane spanning G-protein coupled receptor. A smaller molecule called *retinal* is attached via a Schiff bond to residue Lys 296 in the $7^{th}$ transmembrane domain of the photopigment. In the dark, the side chain of retinal is bent at the $11^{th}$ carbon atom. In this form it is called *11-cis-retinal*. If this molecule absorbs a photon, it undergoes photoisomerization forming straight chain version, known as *all-trans-retinal*. All-trans-retinal unleashes a series of conformational changes in the protein opsin fragment producing metarhodopsin II, which is the activated form of rhodopsin. Most of the conformational changes occur in less than a millisecond, but the last transformation, from metarhodopsin II to metarhodopsin III, requires several minutes to accomplish. Ultimately metarhodopsin III dissociates into opsin and all-trans-retinal. All-trans-retinal is subsequently reduced to vitamin A (all trans-retinol), which is synthesized back into 11-cis-retinal. It re-associates with opsin, completing the cycle (cf. Kingsley, 1996).

Metarhodopsin II initiates the second stage of phototransduction process via activation of an associated $G_t$ molecule known as *transducin*. Transducin is a typical G-protein, composed of α, β and γ subunits, which is activated by exchange of guanosine diphosphate (GDP) for guanosine triphosphate (GTP) within its α-subunit. Upon activation the $G_t$ α-subunit is transferred to and activates a phosphodiesterase (PDE) that hydrolyzes cytoplasmic cyclic guanosine monophosphate (cGMP) to guanosine monophosphate (GMP). Reduction in the concentration of cytoplasmic cGMP in the photoreceptor outer segment releases bound cGMP from the cyclic nucleotide-gated (CNG) ion channels. Dissociation of cGMP from the CNG ion channels initiates the final stage in the phototransduction process, the inactivation of the $Na^+$ currents through these CNG channels in the photoreceptor outer segment. This complex multistage photochemical process might seem cumbersome, yet it affords huge amplification of the incoming signal. One molecule of photopigment can absorb only a single photon. But the resulting metarhodopsin II activates approximately $10^2$ $G_t$ molecules per second, each of which activates approximately $10^3$ PDE molecules per second. Therefore, a single photon can lead to hydrolysis of approximately $10^5$ cGMP molecules per second (cf. Chabre and Deterre, 1989). This biochemical amplification is directly responsible for the remarkable sensitivity of the retina to light. Elaborate experiments have shown that the human is capable of consciously detecting the absorption of a single photon by a rod (cf. Kingsley, 1996).

---

[5] The *plasmalemma* (also called *cell membrane* or *plasma membrane*) is a biological membrane separating the interior of a cell from the outside environment.

[6] The reader should be warned that the *outer segment* of rods and cones actually lies deep in the eye. Thus the light rays pass all retinal layers starting from the ganglion cells, passing through the bipolar cells and ending with the photoreceptors (see *Figure 2* for details).

[7] Alternative naming of opsins: (1) long wavelength sensitive (LWS) (red), (2) middle wavelength sensitive (MWS) (green), (3) short wavelength sensitive (SWS) (blue).

[8] Alternative terminology is L-cones, M-cones and S-cones, where L, M and S refer to long (red), middle (green) and short (blue) wavelengths of detected photons.

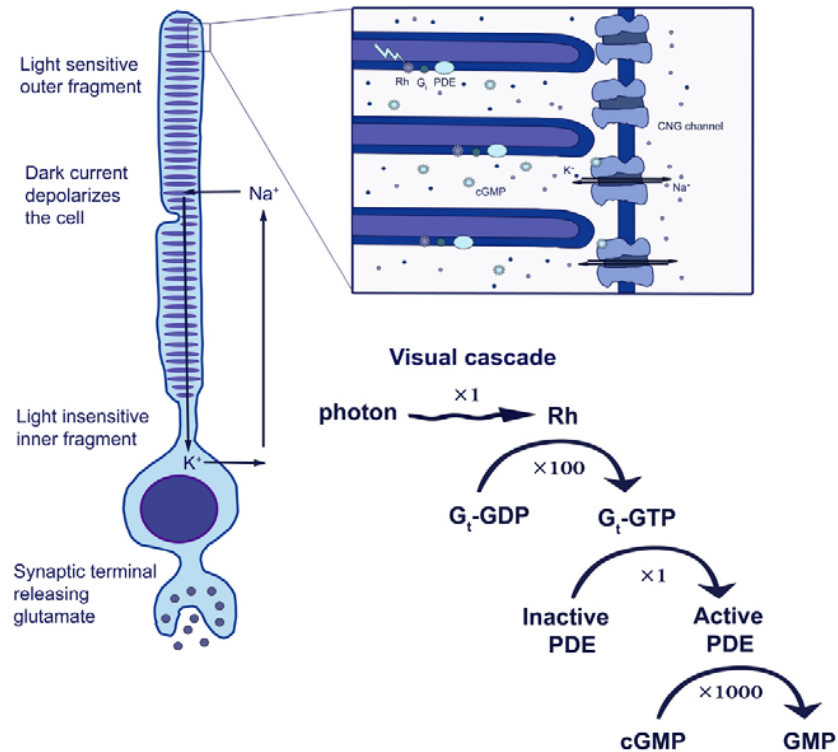

**Figure 1.** Biochemical events in rod phototransduction. Legend: cGMP, cyclic guanosine monophosphate; CNG channel, cyclic nucleotide-gated ion channel; GDP, guanosine diphosphate; GTP, guanosine triphosphate; $G_t$, transducin; PDE, phosphodiesterase; Rh, rhodopsin.

In the dark, photoreceptor cells have a resting potential of about -40 mV. This relatively small value is due to a steady current flow through the plasma membrane, called the *dark current*, which is carried by two ions ($Na^+$ influx and $K^+$ efflux) through CNG ion channels. The CNG channels conduct $Na^+$ and $K^+$ ions almost equally well and consequently have reversal potential around 0 mV (cf. Lu and Ding, 1999). Therefore when opened the CNG channels tend to depolarize the cell.[9] If the photoreceptor cell is illuminated, cytoplasmic cGMP concentration decreases and disrupts the ionic current through the CNG channels, thereby hyperpolarizing the cell. The cell hyperpolarization in effect decreases the release of neurotransmitter (glutamate) at the base of the rods and cones. Thus unlike ordinary neurons, which release transmitter from the synaptic button as a discrete event in response to an action potential, in photoreceptors there is a continuous release of neurotransmitter from the synapses, even in the dark. Modulation (decrease) of the dark current under photon detection serves to modulate (decrease) the release of neurotransmitter from the receptor cells, which means the photoreceptors transmit sign-reversed information. *Thus the first essential feature of the retina is that it amplifies the photon signal and converts it into macroscopic electric currents.*

---

[9] When the CGN channel is open, the resting membrane potential of -40 mV is dragged towards the reversal potential of the CGN channel, which is 0 mV. Thus the photoreceptor is depolarized.

### 2.1.2 Processing of visual information by interneurons

The *retina* of vertebrates shows a layered structure of different neuronal cell types. Retinal neurons can be anatomically classified as (1) photoreceptors, (2) horizontal cells, (3) bipolar cells, (4) amacrine cells and (5) ganglion cells. These different cell types are arranged in ten layers and each cell type has a specific function. The visual information is transduced into electric signal by photoreceptors and is then transferred both horizontally across and vertically through the retina.

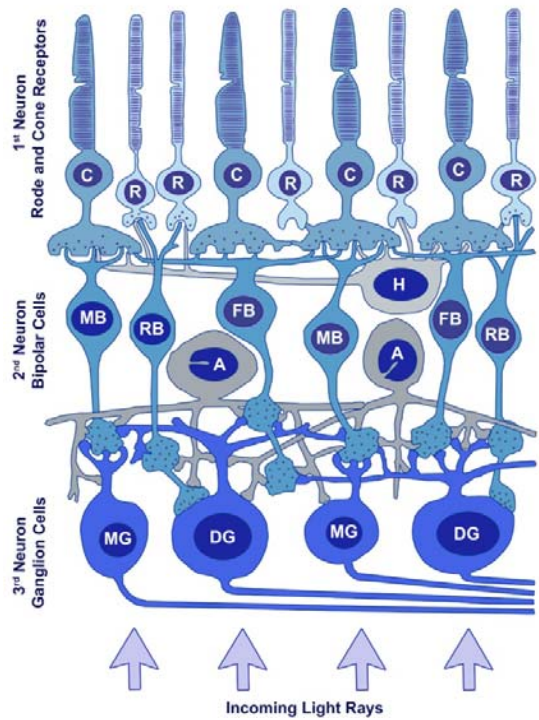

**Figure 2.** Schematic representation of the major cell types in the retina according to Dowling and Boycott (1966). Vertical transfer of information: 1st neuron - rods and cones, 2nd neuron - bipolar cells, 3rd neuron - ganglion cells. Horizontal transfer of information is mediated by interneurons (horizontal and amacrine cells). Legend: C, cone; R, rod; A, amacrine cell; H, horizontal cell; MB, midget bipolar cell; FB, flat bipolar cell; RB, rod bipolar cell; MG, midget ganglion cell; DG, diffuse ganglion cell.

The photoreceptors, the bipolar cells and the ganglion cells are glutamatergic neurons, which transmit the visual information vertically (towards the brain cortex). The vertical organization passes through the bipolar cells in order to reach the ganglion cells. In the retina there are over $10^8$ receptor cells but only about $10^6$ ganglion cells. This means that considerable convergence takes place along the vertical pathway. For example, in the cat *fovea*[10], approximately 200 receptor cells affect a single ganglion cell (cf. Kingsley, 1996). Conventionally only the vertically transmitting neurons are numbered e.g. the photoreceptor rods and cones are referred to as 1st neuron, the bipolar cells are referred to as 2nd neuron, and ganglion cells are referred to as 3rd neuron in the *visual pathway*, etc., however here we would like to pay special attention to the functional role interneurons (for a general plan of the retina see *Figure 2*).

In addition to the convergence along the vertical pathway the horizontal and amacrine cells provide the mechanism for the lateral spread of information horizontally across the retina and ensure that a single receptor cell can affect several adjacent ganglion cells (Kingsley, 1996). Horizontal cells and amacrine cells are *interneurons*, which transmit the information horizontally within the retina. The horizontal cells are GABAergic. In contrast, amacrine cells are in the majority either GABAergic or glycinergic interneurons; however additional neurotransmitters or co-transmitters in amacrine cells have been identified such as acetylcholine and dopamine (cf. Masland 2001; Kolb *et al.*, 2010).

Notably interneurons are responsible for the *lateral inhibition* mechanism, which allows effective detection of "edges" (regions where the light intensity changes abruptly). The *lateral inhibition* mechanism, by which boundaries between dark and bright areas are enhanced, is revealed most strikingly in some visual illusions (see *Figure 3*). Indeed the existent visual illusion due to the lateral inhibition mechanism means that not only the photon signal is amplified before it enters the brain cortex, but also that the visual information is processed (modified and irreversibly changed) before it enters the cortex.

---

[10] The *fovea* (meaning *pit* in Latin) is the center most part of the *macula* (an oval-shaped highly pigmented yellow spot near the center of the retina of the human eye). The fovea is responsible for our central, sharpest vision. In humans it has a very high concentration of cones (allowing us to appreciate color) and notably lacks rods. Also in primates due to the existence of so called midget system, actually a single cone might be associated with a single (midget) bipolar cell and single (midget) ganglion cell.

*Thus second essential feature of the retina is that it irreversibly modifies the visual information and sends processed via lateral inhibition signal to the brain cortex.*

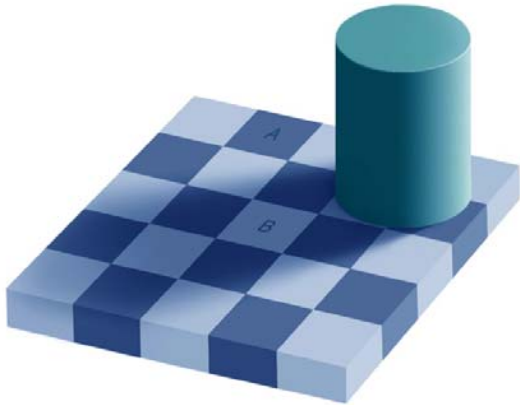

**Figure 3.** An optical illusion. The square A is exactly the same shade and color as square B. The effect is notably due to the lateral inhibition mechanism in the retina, which predicts that square surrounded by lighter squares should look darker due to the lateral inhibition from the surrounding excited lighter squares. One can easily verify that both squares A and B are exactly the same shade and color if deletes everything else from the figure (e.g. using image editing software program).

Up to this point we have explicitly formulated the two most important points (facts) about visual perception that should be kept in mind by any researcher: namely there is (1) amplification and (2) irreversible processing of the visual information in the retina. The following description of how the visual information is converted into digital (binary) signal by the ganglion cells, and delivered through the lateral geniculate nucleus (LGN) (located in the thalamus) to the brain cortex is provided for completeness, but is in the most part unnecessary for showing that the quantum teleportation proposal along the visual pathways is a pseudo-scientific enterprise.

## 2.2 Receptive fields of the ganglion cells

The modulation of neurotransmitter release would be of no benefit if bipolar cells responded to the transmitter with action potentials. All subtle changes in transmitter concentration above and below threshold would be lost in the all-or-nothing response. Therefore, it is not surprising that except for the *ganglion cells*, none of the retinal cells display action potentials. Instead all of the signaling within the retina is performed by graded potentials. Only when the information must be conveyed a considerable distance from the retina is the visual information converted to a *digital form* (action potentials). This function is performed by the ganglion cells (the 3rd neuron in the visual pathway), which project to thalamus.

One can record the action potential from a single ganglion cell in response to white stimuli applied to the retina. They respond to light presented to a restricted locus in the retina, which is the *receptive field* of the ganglion cell.

• One type of ganglion cell is the ON-center ganglion cell, which will respond by increasing its firing rate if one shines a small spot of light on the retina within its receptive field (cf. Westheimer, 2007). If the size of the spot is increased, the ON-center ganglion cell response also increases. If the size of the illuminating spot is increased beyond a certain point, the response of the ganglion cell begins to decrease. Beyond a certain size, further increases in the spot size have no additional effect on ganglion cell output (cf. Kingsley, 1996). These observations suggest that the responses of the ON-center ganglion cells represent the influence of a large number of adjacent receptors. Those in the center of the illuminated field converge on and excite the single ganglion cell. Receptors outside this central core of synergistic receptors inhibit the ganglion cell, since beyond a certain spot size increasing the size of the illumination reduces ON-center ganglion cell output. Physiologically the excitation of ON-center ganglion cells is achieved by ON-center bipolar cells, which express metabotropic inhibitory mGluR6 receptors (Vardi *et al.*, 2000; Duvoisin *et al*., 2005; Tian and Kammermeier, 2006). Since illumination by light of rods and cones leads to decreased release of excitatory neurotransmitter (glutamate), and mGluR6 are inhibitory due to $G\alpha_o$ coupling, the ON-center bipolar cells are thus excited by light due to disinhibition.

• Another type, called OFF-center ganglion cell, has been discovered that has inverse receptive properties. The center must be dark while the surrounded area is illuminated. Experimentally, it has become evident that the pattern of illumination that maximally excites the OFF-center ganglion cell looks like a doughnut (cf. Westheimer, 2007). The doughnut hole is the center while the doughnut itself represents the surround. The OFF-center ganglion cells are excited by OFF-center bipolar cells, which express excitatory ionotropic glutamate receptors such as AMPA and kainate receptors (cf. Kamphuis *et al*., 2003), and this makes perfect sense in view of the sign-reversed transmission performed by rods and cones. A reader who is acquainted with basic neurobiology can easily verify that due to expression of AMPA and kainate receptors the OFF-center bipolar cell and the corresponding OFF-center ganglion cell will be excited when the central photoreceptor is at dark.

Both types of ganglion cell, ON-center and OFF-center cells are about equally represented in the mammalian retina. The most effective size of the central spot varies with its location on the retina. In the cat fovea it is about 30´´ (seconds of a degree), while at the periphery it may be as large as 80´´. By comparison, the size of an individual cone in the fovea is about 4´´ (cf. Kingsley, 1996). In the human fovea due to the existent *midget system* (one cone excites a single *midget bipolar cell* and respectively single *midget ganglion cell*) it is considered that the spatial resolution of the fovea is only limited by the size of the cones (cf. Masland, 2001). The latter claim has been elegantly proved via adaptive optics scanning laser imaging by Rossi and Roorda (2010).

At this point one might well ask what the advantage of representing an image as field of spot annuli as opposed to a field of simple pixels is. The center-surround response characteristic of retinal ganglion cells is simply the visual manifestation of the *principle of lateral inhibition*. The lateral inhibition manifested in retinal center-surround receptive fields enhances the boundary between the light and dark areas of the image, which are subsequently emphasized in the neural signaling of the ganglion cells. Therefore, *the information conveyed to the central visual structures carries information not inherent in the simple pixel representation of the light impinging onto the photoreceptors*.

Lastly, we should mention that even though most ganglion cells respond to the center-surround pattern, they can be subdivided into different functional classes based on their temporal responses to stimuli. In cat the three major functional classes are named X, Y and W cells (cf. Stone and Fukuda, 1974) and correspond respectively to morphologically defined β, α and δ ganglion cells (Saito, 1983).

• The axons of X cells project to the parvocellular cells (laminae 3-6) in the LGN. They have small receptive fields (about 10) and for the most part carry information from the central area of the retina. They respond during the presentation of light, roughly in proportion to its intensity.

• The axons of Y cells project to the magnocellular cells (laminae 1-2) in the LGN. They have larger receptive fields (about 30) and originate, for the most part, in the peripheral portions of the retina. Y cells generally respond at the onset and termination of the stimulus and show a preference for stimuli moving quickly across the visual field.

• The axons of W cell project mostly to the *superior colliculus* and *pretectal area*. W cells have unusual response requirements that do not correspond with the center-surround receptive fields of the X and Y cells. They have very slow conduction velocities, respond to general levels of illumination and are involved in regulating the iris (cf. Kingsley, 1996).

### 2.3 Central visual pathways

Axons of the ganglion cells leaving the *temporal half*[11] of the retina traverse the optic nerve to the optic chiasm, where they join the *optic tract*[12] and project to the *ipsilateral*[13] cerebral hemisphere (see Figure 4). Axons of the ganglion cells leaving the *nasal half* of the retina cross the midline at the chiasm and terminate in the *contralateral*[14] cerebral hemisphere. This arrangement means that all axons in the optic tract carry information about the contralateral visual field. This pattern of hemispheric reversal with respect to the external world is also seen in the organization of motor and other sensory systems[15]. Axons of the optic tract generally terminate in three areas of the central nervous system: (1) the lateral geniculate nucleus (LGN) of thalamus, (2) the superior colliculus and (3) the pretectal area. The trajectory through the LGN is the largest most direct and clinically most important pathway by which visual information reaches the cerebral cortex. A second pathway passes through the superior colliculus and the pulvinar before reaching the cortex. The third pathway from the retina does not reach the cerebral cortex. It terminates in the brain stem pretectal nuclei that regulate eye movements, lens accommodation, and pupillary size.

About 80% of the optic tract axons synapse in the lateral geniculate nucleus (LGN). The LGN is a laminated structure, having 6 layers (laminae). The ipsilateral fibers of the optic nerve terminate in laminae 2, 3 and 5, while the contralateral fibers terminate in laminae 1, 4 and 6 (cf. Hübel, 1995; Kingsley, 1996). The ventromedial laminae 1 and 2 contain large cells and therefore called the *magnocellular division*[16], while the remaining laminae 3-6 have smaller cells and are known as the *parvocellular region*[17]. Ganglion cell axons are distributed in the LGN according to their origin in the retina and thus establish a *retinotopic organization* of LGN (Voigt *et al.*, 1983). Projections from the retina are not distributed to the LGN in proportion to their spatial distribution in the retina, but in proportion to their receptor density. For example the central 20° of the retina occupies about 10% of the retinal area, yet contains about half of the total number of receptors. The ganglion cells that receive receptor input from the central 20° send axons that synapse in about 65% of the LGN (cf. Kingsley, 1996).

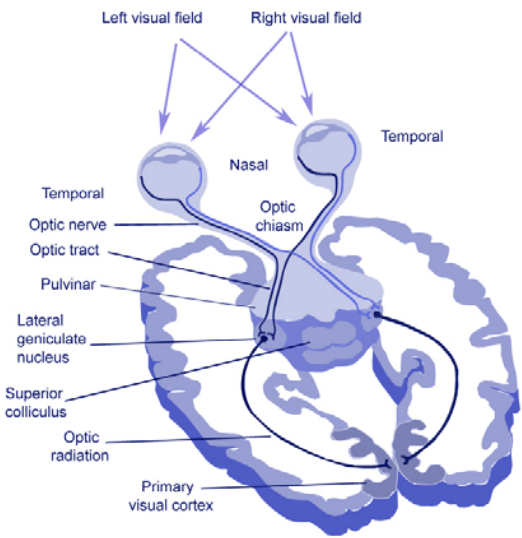

**Figure 4.** The main visual pathway. The optic nerve is composed by the axons of the ganglion cells. After the chiasm most of the axons forming the optic tract terminate in the ipsilateral lateral geniculate nucleus (LGN). The axons of the LGN neurons form the optic radiation and terminate in the striate cortex (V1).

There are about $10^6$ neurons in each LGN, all of which project to the *ipsilateral occipital*[18] *cortex* (area 17) as the *optic radiations*. These axons fan out as they leave the LGN. Some LGN axons take a direct route to the occipital pole, terminating in the superior lip of the calcarine gyri[19], while others take an indirect route through the temporal lobe before reaching the inferior lip of the calcarine gyri. The portion of the cerebral cortex that receives LGN axons is called the *striate cortex* (Brodmann area 17) and is usually labeled as V1 in order to designate it as the *primary visual cortical area*. The expanded neural representation of the fovea found in the retina and LGN is maintained in the visual cortex. The fact that significantly more cortex is dedicated to the processing of information originating from the fovea than other retinal locations is known as *cortical magnification* (Daniel and Whitteridge, 1961).

---

[11] Anatomically the term *temporal* refers to the *temporal bone*, which is a part of the skull and contains part of the ear canal, the middle ear, and the inner ear.

[12] The *optic tract* is a continuation of the optic nerve and runs from the optic chiasm to the lateral geniculate nucleus.

[13] The term *ipsilateral* refers to located on the same side structures.

[14] The term *contralateral* refers to located on the opposite side structures.

[15] With few exceptions, due to crossing of peripheral sensory and motor pathways, the left cerebral hemisphere predominantly senses and controls the right part of the body and vice versa.

[16] The adjective *magnus* means *large* in Latin.

[17] The adjective *parvus* means *small* in Latin.

[18] The adjective *occipital* refers to the *occipital bone,* which forms the back part of the skull and the base of the cranium.

Most of the remaining axons of the optic tract terminate subcortically in the *superior colliculus,* which is a region of the midbrain regulating eye motion. The superficial layer of the superior colliculus receives both retinal and cortical visual information, the latter descending from V1. The latter fact is not surprising provided that eye motion is not subject solely to non-conscious reflexes, but is also under direct conscious control. Neurons in the superficial superior colliculus project both to the *pretectal area* (responsible for the pupillary light reflex) and the *pulvinar*, which is a large nucleus of the thalamus responsible for focusing the attention on salient features of the visual image (cf. Grieve *et al.*, 2000). In primates, the pulvinar neurons do not synapse in V1, instead they project to a number of extrastriate cortical areas, including Brodmann areas 18 and 19 and the temporal lobe, particularly the superior temporal gyrus. These regions are important for processing highly abstracted visual perceptions (cf. Kingsley, 1996).

## 2.4 Central processing of visual information

The thalamocortical projections from LGN terminate in layer IV of the *primary visual cortex* (area V1). The *primary visual cortex* is mainly buried in the *calcarine fissure*[20]. Almost all information in the visual system is recognized as being processed by V1 first, and then passed out to higher order systems (cf. Zeki, 1993). Electrophysiological experiments have established that there is a precise map of how segments of the retina are related to areas of V1. V1 has a *retinotopic map* in that one spot in the visual field maps directly to a spot on V1 (Schwarz, 1980a, b). The overall details of this map are well known from animal studies or functional MRI and PET studies with humans (Holmes, 1945; Fox *et al.*, 1987; Engel *et al.*, 1997). The upper visual cortex receives signals from the lower visual field and similarly, lower visual cortex processes information from the upper visual field. The right visual cortex processes the left field of view and vice versa. Central vision is represented at the pole of occipital cortex. This general representation is referred to as *cruciform model* (cf. Ahlfors *et al.*, 1992). Within the retinotopically organized V1 cortex the cortical area devoted to process information per degree of visual angle is larger for central vision compared to peripheral vision. This implies that the scale of the retinotopic mapping changes as a function of retinal location. The term *cortical magnification* introduced by Daniel and Whitteridge (1961) is used to describe this phenomenon and refers to the fact that a very large number of cortical neurons process information from a small region of the center of the visual field corresponding to the retinal fovea. If the same stimulus is seen in the periphery of the visual field (i.e. away from the center), it would be processed by a much smaller number of cortical neurons (which will have respectively smaller cortical surface area).

### 2.4.1 Receptive fields of the cortical visual neurons

One can study the properties of cortical cells in the visual system in same manner as one studies retinal ganglion cells, by presenting visual stimuli to the retina and recording the activity of the cortical cells. Just as with ganglion cells, one can determine the most effective stimulus shape and map their receptive fields. The most effective stimulus shape for individual neurons in V1 is different from the *center-surround* shape that is so effective in driving ganglion cells in the retina and LGN. In the visual cortex, neurons respond best to rectangular shapes (cf. Hübel, 1995; Kingsley, 1996). Neurons in V1 also respond in accordance to other criteria that allowed them to be categorized into 2 broad classes: *simple cells* and *complex cells* (Hübel and Wiesel, 1959; 1962; 1968).

---

[19] *Gyri* is the plural of *gyrus*, which is a ridge on the cerebral cortex.
[20] Calcarine fissure is a deep sulcus situated on the medial surface of the *occipital cortex*.

In order to understand how simple and complex cells operate in primary visual cortex V1, first one should be acquainted with some of the elementary functions (gates) in *Boolean logic*. The fact that digitally represented signals enter and leave each neuron allows one to use Boolean concepts in analyzing at least some features of the neural networks and liken them to digital computers. Modern digital computers are based on digital electronic circuits within which the signals interact at gates that perform specific Boolean functions. All Boolean functions could be formed by various combinations of 3 *elementary functions*: AND, OR and NOT[21]. These functions could be best appreciated by building a *truth-table* that illustrates all of the input-output relationships of a gate (cf. Mendelson, 1970; Kingsley, 1996).

Table 1. Truth-tables of elementary Boolean gates.

| NOT | | AND | | | OR | | |
|---|---|---|---|---|---|---|---|
| Input | Output | Input 1 | Input 2 | Output | Input 1 | Input 2 | Output |
| 1 | 0 | 1 | 1 | 1 | 1 | 1 | 1 |
| 0 | 1 | 0 | 1 | 0 | 0 | 1 | 1 |
| | | 1 | 0 | 0 | 1 | 0 | 1 |
| | | 0 | 0 | 0 | 0 | 0 | 0 |

• A simple cell in V1 responds best to a bar of light oriented at a particular angle in the visual field (Hübel, 1995). The bars must be presented within receptive fields that are also rectangular. The cells respond best to stimuli having excitatory centers and inhibitory surrounds or vice versa. The effective stimuli and receptive fields for simple cortical cells are entirely analogous to the spot-annulus fields of the ganglion cells except for their shapes. One attractive explanation for the bar-shaped receptive fields of simple cells supposes that several geniculate neurons with spot-annular receptive fields that overlap in the shape of a bar converge on a single cortical cell. The simultaneous activity of all this converging geniculate cells would be required to cause the cortical cell to fire. This is the neural equivalent of a logical AND function.

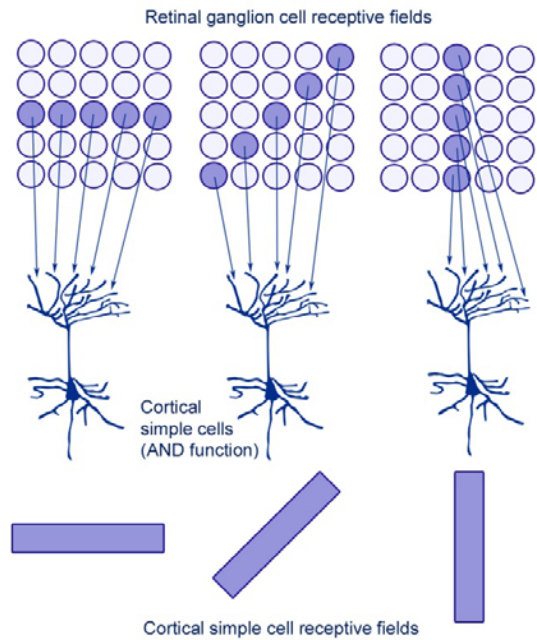

**Figure 5.** Simple cells in V1 have rectangular receptive fields. The response of each simple cell to light varies according to how well the stimulus matches the rectangular receptive field. The receptive fields of LGN cells (circles) are represented with connections going to several simple cortical cells in V1. The simple cell receptive properties are modeled by a Boolean AND gate relating the output of several LGN cells.

• A complex cell in V1 has much larger response field compared to simple cell. The complex cell also responds best to a properly oriented bar of light, but the bar may appear anywhere in the large receptive field of the cell (cf. Hübel, 1995). In other words, complex cells respond to an abstracted version of the stimulus that excites simple cells. The quality that is abstracted is location within the visual space. The behavior of complex cells can be explained by assuming that all the simple cells of the same orientation but with receptive fields in different parts of the visual space synapse with a single complex cell and that each simple cell is capable of causing the complex cell to fire. This is the neural equivalent of a logical OR function.

**Cortical simple cell receptive fields**

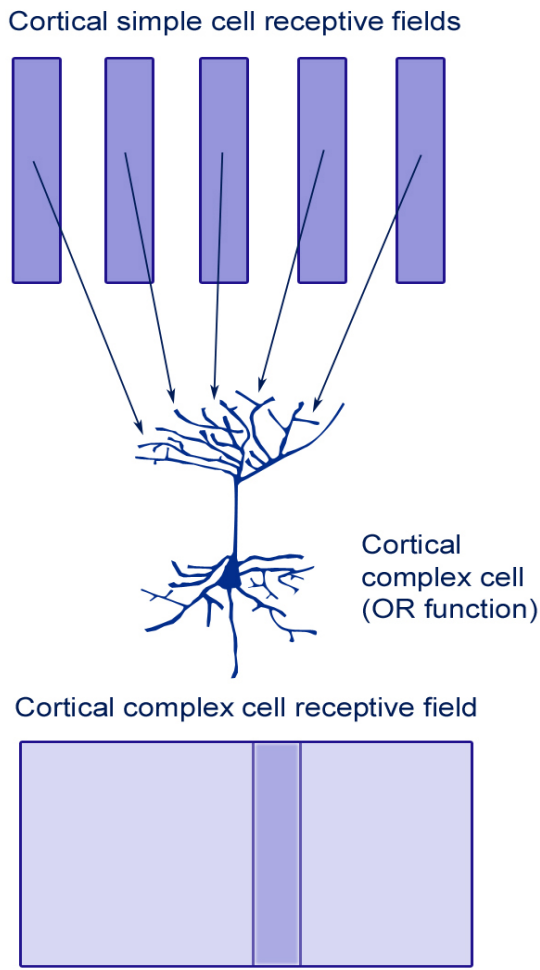

Cortical complex cell (OR function)

**Cortical complex cell receptive field**

**Figure 6.** Complex cells in V1 have larger receptive fields. The synaptic drive for complex cells is thought to be derived from the output of many cortical simple cells, where simple cell receptive fields are depicted as bars within the complex cell receptive field. If any simple cell with appropriate response characteristics in the receptive field of the complex cell is able to drive the complex cell firing, this would correspond to a Boolean OR gate.

### 2.4.2 Columnar organization of the visual cortex

Cortical regions as a rule are organized into *functional columns*. In the primary visual cortex, there are two types of functional columns depending on the origin of the LGN thalamic afferent axons[22], which terminate in cortical layer IV. The cells in one functional column receive afferents only from LGN laminae 2, 3 and 5, which bring information from the *ipsilateral* eye. An immediately adjacent column receives afferents from LGN laminae 1, 4 and 6, which bring information from the *contralateral eye*. The two types of functional columns are called *ocular dominance columns*. Thus the origin of the visual information alternates within cortical layer IV from the left to the right eye, across functional columns. A pair of left-right ocular dominance columns corresponding to the same spot in the visual field is called a *hypercolumn*. Hypercolumns are organized across V1 according to the retinotopic map that represents the visual fields (cf. Kingsley, 1996). Each hypercolumn is subdivided into large number of *orientation columns*. All simple cells within each orientation column respond to bars of light of the same orientation in the visual field. Across a hypercolumn, different possible angles are represented in separate orientation columns (*Figure 7*).

Sprinkled among the hypercolumns are *blobs*, a term used to describe a cortical column that processes color information and does not have orientation requirements[23]. The cerebral cortex is also horizontally organized. Pyramidal cells within layers III and V send axons horizontally across the cortex for several millimeters. These axons send vertical arborizations at intervals that correspond to the width of hypercolumns. Color blobs and orientation columns of the same angle are interconnected across hypercolumns, which accounts for the observed interactions between the left and right ocular dominance columns in *binocular vision* (cf. Kingsley, 1996).

---

[22] *Afferents* refers to nerve fibers (axons) that input information to the cerebral cortex (or other central processing unit). In contrast, *efferents* refers to nerve fibers (axons) that output information from the cerebral cortex (or other central processing unit).

[23] *Blobs* are named for the blob shape of the axonal arborizations of the thalamic afferents that arise from intralaminar neurons in the LGN and terminate in layers II and III of the cortex.

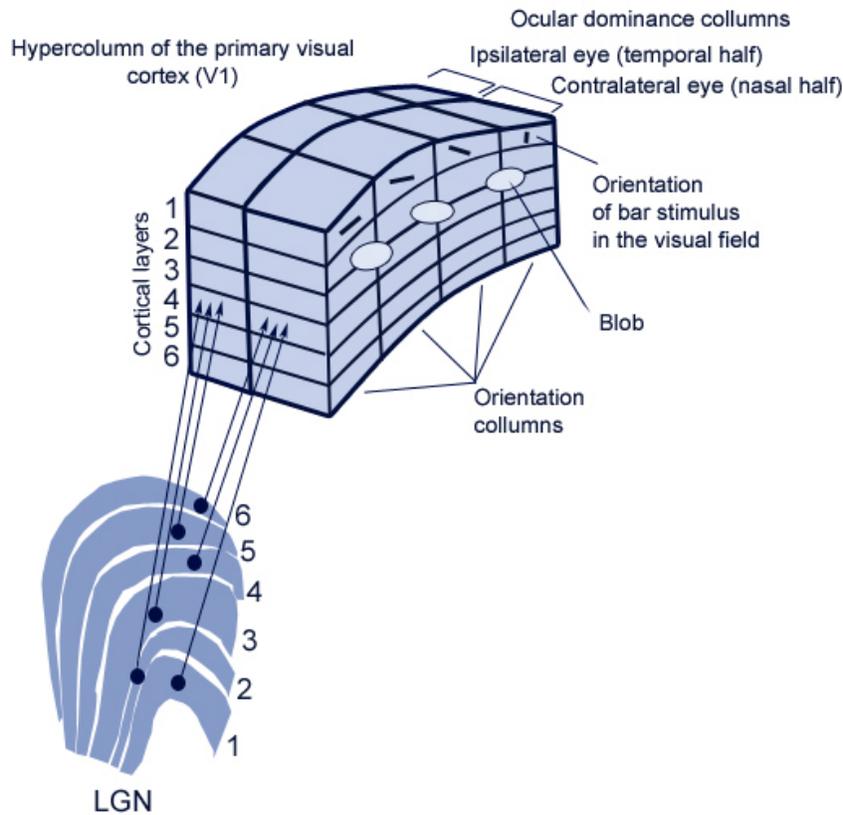

**Figure 7.** Hypercolumns in the visual cortex consists of pairs of ocular dominance columns and numerous orientation columns. Interspersed in the hypercolumns are blobs - columns that are not orientation sensitive and appear to process color information. Each hypercolumn processes information from a single discrete spot of the visual field. Hypercolumns are arranged across the surface of the primary visual cortex in a retinotopic pattern.

### 2.4.3 Parallel processing in the visual system

The visual system processes the information along separate *parallel pathways*. There are four parallel channels of information: one scotopic (monochromatic) for the rods and 3 photopic (red, green and blue) for the cones. The parvocellular (X cell) and magnocellular (Y cell) pathways divide the visual space into a high-acuity path (X) with static properties and low-acuity path (Y) that is very sensitive to movement. In addition, this visual information is segregated spatially according to position within the visual field. All of these channels are segregated at the level of the retina (cf. Kingsley, 1996).

Further divisions occur at the level of thalamus. Visual information passing through the LGN is directed to V1 of the cerebral cortex. An alternative path through the *pulvinar*[24] innervates all other visual areas in the cortex except V1. Within the LGN and cerebral cortex the parvocellular and magnocellular channels remain segregated. The parvocellular path synapses in the deepest part of layer IV, while the magnocellular path synapses more superficially in layer IV. The parallel distribution of information is elaborated further by the output pattern from the cortical columns. The upper layers IV and II of the cortex project to other cortical areas by intracortical association fibers. The deeper layers project to the *superior colliculus*, the *pulvinar* (from layer V), and *LGN* (layer VI) (cf. Kingsley, 1996).

---

[24] *Pulvinar* is one of the thalamic nuclei (see Figure 2).

### 2.4.4 Poststriate processing

Feature abstraction continues beyond V1 in the extrastriate visual areas. As many as 20 different retinotopically mapped areas have been discovered (cf. Kingsley, 1996). In general terms the anatomical organization of these areas utilizes both parallel and serial schemes for processing of visual information. At least 2 parallel information channels leave V1. The magnocellular (Y) pathway appears to extract information regarding *motion*, interocular disparity (which is necessary for depth perception), and spatial relationships. It proceeds serially through several individually mapped regions into the *posterior parietal cortex*. The parvocellular (X) pathway analyzes *form*, color, and interocular disparity. It projects in serial stepwise fashion to the *temporal cortex*. These two pathways are known as *dorsal occipitoparietal "where" pathway* and *ventral occipitotemporal "what" pathway* (Ungerleider and Pessoa, 2008; Karnath *et al.*, 2009). The *blob* system, which is a subset of the parvocellular system, seems to be exclusively associated with color analysis and may represent a third parallel path. These parallel pathways are not strictly separated and interact at several levels.

The various extrastriate visual areas seem to individually abstract certain global attributes from the visual image such as shape, color, motion, and interocular disparity. Moreover, clinical data have shown that abstractions of the visual scene are analyzed and individually brought to conscious perception by anatomically separate part of the cerebral cortex. Indeed appropriately placed cortical lesion would produce perceptual losses of specific attributes while other attributes would be preserved (cf. Kingsley, 1996).

Lesions that produce a defect in recognition or meaning without losses in objective sensation are termed *agnosias*. Agnosia is an inability to recognize or attach meaning to stimuli in cases where the input of sensory information is not impaired and there is no generalized intellectual loss (Farah, 1992). Thus agnosia is not a sensory loss in the sense that *blindness* or *deafness* is sensory loss.

Several visual agnosias have been correlated with specific cortical lesions. For example, *object agnosia* is the inability to recognize familiar objects by visual means alone. The objects might be readily recognized by tactile manipulation, auditory cues or odor, but visual sensation itself does not make the object recognizable or meaningful to the patient. Visual recognition is not possible because the visual sensations, even though consciously perceived hold no meaning to the patient. The fact that there is no impairment of visual sensory input has been rigorously proved by a case study of object agnosia in an artist following infarction, who retained various techniques (perspective, shadowing, designation of texture) which allowed him to copy displayed objects in a veridical fashion (Wapner *et al.*, 1978). Other curious cases such as visual agnosia for line drawings and silhouettes without apparent impairment of real-object recognition have been also reported (Hiraoka *et al.*, 2009). Object agnosias are associated with bilateral lesions to ventromedial portion of the occipitotemporal cortex (Brodmann areas 20 and 21).

*Prosopagnosia* is an extreme form of object agnosia manifested as an inability to recognize faces. Such patients cannot recognize people from their face, but have no difficulty recognizing them by their voice, clothes or body movement. Face-specific neurons have been found in the inferotemporal cortex of the monkey (Bruce *et al.*, 1981; Perrett *et al.*, 1982; Eifuku *et al.*, 2004). Within it, single cells have been discovered that respond only to visual stimuli in a shape of a simian hand or the outline of a simian face. Such cells are encountered rarely and have been dubbed "grandmother cells", i.e. neurons that may be so specific as to respond only to one's grandmother (cf. Jagadeesh, 2009). The discovery of the "grandmother cells" has provided a physiological explanation for prosopagnosia.

*Achromatopsia* is the inability to make color hue discrimination and is associated with bilateral lesions to area 37 (and possibly area 18), which is a transition zone between the occipital and temporal lobes. The patients remark that everything looks gray (cf. Kingsley, 1996). Achromatopsia is a type of agnosia and should not to be confused with *color* *blindness*, which is a retinal defect due to absence of one or more sets of color-sensitive cones. In achromatopsia the color information is sent to the brain, however it has no meaning to the patient.

### 2.5  Bionic eye helps blind man to see

A direct experimental proof that (1) *it is the brain cortex that sees the visual images*, and that (2) *the visual images enter the cerebral cortex in the form of electric signals*, is the bionic eye implant surgery for the blind (cf. Dobelle, 2000). Dobelle has helped a blind man (identified only as Jerry) to see again using a tiny television camera (mounted on a pair of glasses) that is connected to electrodes implanted into the visual cortex of the patient. With the input from the bionic eye the patient can make out the outlines of objects, large letters and numbers on a contrasting background, and can use the direct digital input to operate a computer. Jerry has been blind since age 36 after a blow to the head. In 1978 he got a brain implant consisting of an array of 68 small platinum electrodes attached on the surface of Jerry's brain. The electrodes could be stimulated through wires entering his skull behind his right ear.

With the development of sufficiently powerful computers in 2000 it became possible to connect camera (the bionic eye) to a small computer, carried on Jerry's hip. The computer then highlights the edges between light and dark areas in the camera image and sends the signal to the platinum electrodes. The electrodes stimulate the neurons in the visual cortex, making Jerry perceive dots of light, which are called *phosphenes*. Jerry gets white phosphenes on a black background and has the equivalent of 20/400 vision[25]. If he is walking down a hall, the doorway appears as a white frame on a dark background. Jerry demonstrated by walking across a room to pull a woolly hat off a wall where it had been taped, took a few steps to a mannequin and correctly put the hat on its head. A reproduction of what Jerry sees showed crosses on a video screen that changed from black to white when the edge of an object passed behind them on the screen. Jerry can read two-inch tall letters at a distance of five feet and can use a computer thanks to some input from his son. Although the relatively small electrode array produces tunnel vision Jerry is also able to navigate in unfamiliar environments including the New York City subway system. Jerry says that he can *"see dots of light"*, which is direct evidence that it is the visual cortex that is responsible for the conscious perception of visual information[26]. Moreover, the signal inputted to Jerry's cortex is in the form of electric stimulation, which is consistent with the described *transduction of light* into electric signals that occurs in the retina. Therefore we are to conclude that the visual information entering the visual cortex is encoded in the electric firing of the afferent LGN axons that reach V1.

## 3.  Quantum mechanics and visual perception

### 3.1  Photon echo experiments at Starlab

#### 3.1.1  General description

Saint Hilaire *et al.* (2002) at Starlab devised an experimental system based upon a technique from quantum optics called "photon echo", in order to look for evidence of quantum coherent superposition in the human retina in awake volunteers. In "photon echo" a short laser pulse is sent to the system being studied followed by another pulse from the same source. If *quantum coherence* is occurring in the system a delayed "photon echo" should be detected. At present lasers are used commonly in ophthalmology and the authors claimed the safety of the technique will be assured in the experiments proposed. Because the laser pulses required are of a very low power there would be no risk of injury to the retina and the subjects should perceive only a faint flash of light.

According to Saint Hilaire *et al.* (2002) the Starlab experiment was designed in order to answer the question where does the measurement of a visual field happen: is it right at the first retinal layer, or it happens further up in the visual cortex? The idea of the setup is that the first pulse when reaching the retina causes rhodopsin molecules (and perhaps others) in the rod and cone cells in the retina to go into a state of quantum coherent superposition. If these quantum coherent states persist for a significant time duration (the decoherence time) longer than the interval between the two laser pulses, then the second pulse would cause some of the atoms to precess back towards their initial state where the first pulse was encountered and then emit a photon corresponding to the initial quantum transition (hence the name "photon echo"). This photon echo can be detected with sophisticated quantum optical equipment, and would indicate a state of quantum coherent superposition in the retina (and the brain). According to Saint Hilaire *et al.* (2002) decoherence times for the quantum state in the nanosecond range or higher would be supportive of quantum consciousness theory.

---

[25] This means that Jerry would have to stand 20 feet (6.1 m) from an object to see it with the same degree of clarity as a normally sighted person could from 400 feet (121.9 m).

[26] Another patient who tried the system developed by Dobelle could not "see" anything with it. According to Dobelle the man (aged 65) was blinded at the age of 5 and his brain may have "forgotten" how to use its *visual cortex*. Also it is not known whether the system will work at all for people who were born blind.

The authors supposed that the retina and the visual pathways can exhibit quantum coherence over macroscopic time intervals (e.g. milliseconds). The structures involved in macroscopic quantum superpositions within the retina are claimed to be exactly the *rhodopsin* complexes. After detection of a photon the macroscopic quantum coherent state should be *somehow* preserved through the neural processing, or in other words the photons are expected to collapse not in the retina, but in the brain cortex. Indeed the latter statement is one of 20 testable predictions offered by Hameroff (1998), according to which microtubule-based cilia/centriole structures are quantum optical devices: *"microtubule-based cilia in rods and cones directly detect visual photons"*. In its essence the idea is that microtubules can act as *optical waveguides* to transmit the photons to the cerebral cortex. Despite the fact that so far there are no published results from such "photon echo" experiments, here we will provide brief explanation why this experiment cannot have the claimed nanosecond or longer decoherence times.

### 3.1.2 Evidence against

There is both theoretical and experimental evidence against the proposal by Hameroff (1998) and Saint Hilaire *et al.* (2002). As a first point, if one assumes that microtubules detect directly photons in the retina, it will be impossible to explain the *color blindness*, which results from mutations in the red, green or blue opsin genes. Moreover it is surprising, why only the microtubules in rods and cones detect the visual photons, and not the microtubules in the ganglion cells. After all the ganglion cells are amongst the first retinal cells through which the incoming light rays are passing (see *Figure 2*).[27]

A second, much more severe point is that although the normal irradiance of the retina by the sun is approximately 0.1 mW cm$^{-2}$ under daylight conditions (Clarkson, 1989), if the perceived by the retina amount of light is going to be transmitted by microtubule waveguides towards the brain cortex it would make the optic nerve glow visible light with intensity of approximately 0.1 mW cm$^{-2}$ in the case when the optical nerve is cut[28]. *This is a straightforward prediction and one might easily verify it by cutting an optical waveguide cable. If at one end there is a source of light, this light would be glowing out from the other cut end of the optical waveguide.* Needless to say that if cut the optic nerve is not glowing light. Here we would like to add that possible losses of light on its way towards the retina are irrelevant for the argument, what is discussed here are the possible losses of light within the optic nerve and the efficiency of light transmission by the optic nerve. *If the optic nerve cannot serve as an optic cable then it cannot transmit the visual photons towards the cortex.* A direct way to prove that the optic nerve is not an optic cable is to cut a piece of it and shine light on one of its ends and see if something is coming out from the other end. A desperate attempt to avoid this argument is to maintain that the "cutting" of the optic nerve immediately destroys its properties as an optic waveguide.

---

[27] Curiously, inherently photosensitve ganglion cells do exist, but they are few, express melanopsin, and project to hypothalamus and other subcortical nuclei.

[28] Here we conservatively ignore possible condensing of light due to the fact that the retinal surface receiving the light is greater than the cross-section of the optic nerve.

How this happens and why this should be the case should be answered by Hameroff and his collaborators, who proposed the model at first place.

### 3.2 Gao's approach for distinguishing consciousness
### 3.2.1 General description

According to Gao (2003) the existence of consciousness can be tested using a quantum method. It is a fact that standard quantum theory does not permit non-orthogonal single states to be distinguished. However according to Gao (2003) within "revised quantum dynamics" if the physical measuring device is replaced by a conscious being the non-orthogonal single states could be distinguished. Thus the possible distinguishability of non-orthogonal single states is claimed to provide a physical method to test the existence of consciousness.

Complete mathematical sketch of Gao's argument is provided in *Appendix II*. In essence Gao suggests that the human observer endowed with consciousness is able to evolve through quantum superpositions of states and at the same time the human observer is able to measure (or feel) the dynamical timescale of the collapse of each of these quantum superpositions. Then Gao (2003) proposes that if two non-orthogonal quantum states of a photon are inputted to an observer and if only the second photon state leads to superposition of the observer, then it is possible for the human to understand this fact by feeling longer time passage before the dynamical collapse of the superposed state into one of two observed possibilities.

### 3.2.2 Evidence against

In previous sections of the paper we have shown that introducing a superposition state into the consciousness through the eye is in principle impossible. *This is so because the inputted photons in the retina collapse there and entangle with the amplification biochemical cascade in the photoreceptors, which transforms the inputted visual signal into macroscopic electric currents in the retina*. In other words, amplification of the signal means "wavefunction collapse" or "decoherence". It should be noted that this important fact has been pointed also in previous work (Georgiev, 2002; Thaheld, 2008). The amplification leads to "collapse" (or "decoherence" in no-collapse interpretations of QM) due to the "no cloning theorem" (see detailed proof of this theorem in Wootters and Zurek, 1982). In contrast to classical physics where one is allowed to make any number of copies of a given bit without affecting the bit itself, in QM it is impossible for one to make even a single copy of a given qubit without modifying the qubit. Instead what can be done is creation of qubit copies that are entangled with the original qubit. Such an entanglement with other qubit copies leads to decoherence of the original qubit (mathematically expressed as diagonalization of its density matrix) and is equivalent to wavefunction collapse for those QM interpretations in which the wavefunction collapse is considered to be a real phenomenon. A caveat however is necessary. For no-collapse interpretations of QM there is no such process as wavefunction collapse at all, instead there is only decoherence. Therefore the argument that from photon amplification follows collapse (cf. Thaheld, 2008) is incomplete and cannot be used against no-collapse QM theories such as many-worlds interpretation of QM. First, in no-collapse QM theories it is inappropriate to talk about collapse at all. Second, in such theories one can always perform in principle operation called "quantum erasure". If the "quantum erasure" is applied at a later stage after the amplification, then one can remove all the entangled qubit copies and "undo" the amplification. It would be completely absurd for someone to maintain that the retina amplifies the photon signals and the brain after that uses quantum erasure to "undo" this amplification. However, most of the Q-mind advocates criticized in the current work do not abhor invoking absurd-looking mechanisms just to defend their favorite Q-mind theories. That is why in the current work in order to make the amplification argument solid against possible immunization with "quantum erasure" speculations, we stress explicitly on the processing of the amplified photon signals by the retinal neurons, which leads to some *visual illusions* as a side-effect.

After the photon signal is amplified, processing of the sensory information at the level of neural membranes occurs (e.g. *lateral inhibition* mechanism for detecting "edges" and sharpening the contrast of the input visual image) and the visual information is irreversibly modified before it enters the cerebral cortex (a particular proof for that is the visual illusion shown in *Figure 3*). Exactly because one cannot deny the existence of the visual illusion shown in Figure 3, it is impossible to claim that "quantum erasure" (performed after the visual photons have been amplified) has somehow restored the original coherent state of each of the visual photons. The visual illusion clearly shows that what is experienced by the mind is not identical to the photons that enter the retina. In our opinion, the visual illusions due to the lateral inhibition mechanism in the retina are more sensible and more powerful evidence for the "collapse" of the visual photons compared to mathematical calculations, which might not be immediately clear for most neuroscientists or philosophers (who as a rule do not have any training in quantum physics). Since the discussed (1) amplification of incoming sensory signal, and (2) processing of the amplified sensory signal via lateral inhibition, are fundamental for all human sensory organs, *one cannot use any of the senses in order to input quantum coherently superposed information to the brain cortex* (what could be inputted is a collapsed state or incoherently superposed information, i.e. a state which is *improper mixture*).

Next, we would like to point out that Gao's argument does not rely on the possibility of conscious experience at all, but on the possibility of the system to measure the dynamic timescale of its own collapse. In this case the test system needs to be equipped with a clock (or be a clock), which could perform such a measurement and after reading the clock the system can easily deduce what the measured non-orthogonal states were. Tautologically it follows that according to Gao the consciousness is equivalent with having (or being) a clock, which is able to measure precisely the dynamic timescale of its own collapse.

A third misconception in Gao's approach is concerning the mechanisms of time perception by humans (see *Appendix II* for Gao's requirements for the duration of different collapse times). In previous paper we have discussed available experimental evidence from psychophysical experiments, which show that humans can only discriminate time intervals separated by approximately 40 ms (cf. Georgiev, 2004). The human cortex understands the time passed (and creates subjective feeling of a time flow) only via reading the electric impulses coming from the basal ganglia, which play the role of an internal for the brain clock that measures the objective passage of time. If two events are separated by time interval less than 40 ms, they are perceived as *simultaneous*. Therefore there is no human being that can satisfy Gao's conditions for time perception, since the brain decoheres in a submillisecond timescale far shorter than 40 ms.

### 3.3  Quantum teleportation of photons
### 3.3.1  General description
The last proposal of "quantum" visual perception to be discussed is proposed by Tuszynski's group (cf. Salari *et al*., 2008; Rahnama *et al*., 2009) and involves quantum teleportation of photons. The authors propose that the visual photons are quantum teleported to the brain cortex and they collapse in the cortex instead inside the retina. In order to proceed with further discussion on the feasibility of such an idea, the reader must understand in detail what quantum teleportation is. The mathematics of the teleportation protocol is exhaustively presented in the *Appendix*; here we will outline only the essential steps and requirements.

Suppose that Alice wants to teleport a photon $A$ to Bob. In order to do so, Alice needs another maximally entangled photon pair $B+C$ in Bell basis, with one of the entangled photons $C$ present at the input gate for the teleportation (i.e. present at the location where Alice is), and the second entangled photon from the pair $B$ located at the output gate, where the photon $A$ is to be teleported (i.e. present at the location where Bob is). Here we stress explicitly on the fact that if Alice wants to teleport a qubit $A$, which is a photon, then the necessary maximally entangled qubits *B* and *C* must be identical qubits, or in other words also photons. Next, Alice performs Bell type measurement upon both photons *A* and *C*, which will transform the photon *B* in one of four states, which depend upon the result obtained by Alice from her measurement. If Alice is able to communicate her measurement result to Bob via classical communication channel, Bob can perform a unitary transformation and convert the photon *B* into the original state in which photon *A* was. However without knowing the result obtained by Alice, Bob cannot accomplish the teleportation.

### 3.3.2 Evidence against

Critical moment misunderstood and neglected by Tuszynski's group (cf. Salari *et al.*, 2008; Rahnama *et al.*, 2009) is the fact that if one wants to teleport photons of visible light, he needs entangled photon pair, also of visible light. In other words, *one cannot use quantum entangled microtubules to teleport a photon* (see *Appendix* for detailed mathematics). If one has a source of entangled photon pairs of visible light, which propagate within the optic nerve, then one arrives at the optical waveguide problem discussed in *Section 3.1.2*. Namely, the optic nerve should be glowing light in the case when the optical nerve is surgically cut. The source of light is the one that produces the entangled photon pairs needed for the quantum channel.

Second concern is the fact that the detection of visual photons by retinal photoreceptors leads to *amplification* of the signal via biochemical cascade. A caveat is necessary - the amplified signal is not photons, but an electric current and release of neurotransmitter (glutamate)! This means that the amplified photons are already collapsed, or in other words they are already entangled with other biochemical processes in the retina (such as release of glutamate or not). Even if the teleportation of the state of already amplified photon were possible, then only *entanglement swapping*[29] would occur. This means that the teleported photon to the brain cortex would be also collapsed

(decohered, or entangled with the amplifying biochemical cascades). Teleportation in which entanglement swapping occurs cannot lead to photon collapse in the brain cortex. Simply the collapsed in the retina photon is teleported to the brain cortex in its collapsed state.

Next, we would like to note that the authors make confusion between (1) *tunneling photons* proposed by Mari Jibu and Kunio Yasue (cf. Jibu *et al.*, 1994; Jibu and Yasue, 1995) and (2) *visual photons*. In the first case the tunneling photons are virtual particles that are intimately associated with the motion of water dipoles (forming solitons called *soft polaritons*). In the second case the visual photons are just a form of classical electromagnetic radiation, which if present within the optic nerve could be detected as glowing. In addition, the tunneling photons proposed by Jibu and Yasue are in the *infrared* not the *visual spectrum*. The confusion goes further in supposing that the microtubules first detect the visual photons in a form of dipole displacement of the tubulin dipoles and then teleport it. Needless to say, that this is nothing but a measurement of the visual photon and is thus associated with photon collapse right at the retina. Such a measurement precludes further teleportation of the visual photon to the brain cortex. Particularly if one were able to extract the full quantum information concerning the incoming visual photon from the tubulin dipole displacement alone, then one could create any number of identical photon copies and thus violate the *quantum no cloning theorem* (Wootters and Zurek, 1982). Another technical concern is to further show how the retinal neurons could measure in Bell basis the tubulin dipole displacement in microtubules, and how this will result in action potentials that could be transmitted to the cortex. The latter Bell type measurement is necessary in order to complete the quantum teleportation protocol via unitary transformation at the cerebral cortex (see *Appendix* for mathematical details).

Last, but not least, the neurobiology of vision is completely messed up. For example, the discussion of ipsilateral and contralateral pathways from the retina to LGN in Salari *et al.* (2008), and how they provide means for extracting phase information that is inputted from LGN to the cortex, is a gross misunderstanding. The reader should remind *Section 2.3*, where we have shown that ipsilateral and contralateral pathways remain segregated both at LGN and V1. Namely, LGN laminae 2, 3 and 5, which bring information from the *ipsilateral* eye, project only to ocular dominance columns in the *ipsilateral* V1 (see *Figure 7*). An immediately adjacent ocular dominance column in V1 receives afferents from *ipsilateral* LGN laminae 1, 4 and 6, which bring information from the *contralateral eye*. Thus in LGN there is no neuron upon which the information from both eyes converges[30]. Moreover, the receptive fields of the V1 neurons are bar shaped and the processing of ocular disparity actually happens in V1 and subsequent visual areas. Lastly, we again refer the reader to the optical illusion presented in *Figure 3*, which clearly shows the visual image is not transmitted pixel by pixel; hence individual photons could not be teleported.

---

[29] For detailed exposition on *entanglement swapping* see Grudka *et al.* (2008).

### 3.4 Teleportation of tubulin states

Mavromatos *et al.* (2002) define teleportation as the complete transfer of the coherent state of a microtubule without any direct transfer of mass or energy[31]. This means that the "receiver" microtubule finds itself in an identical state to the "sender" microtubule. Mavromatos and co-workers (2002) demonstrate that given the possibility for entangled states, teleportation between microtubule *A* and microtubule *C* could happen if one follows the protocol described in detail in the *Appendix*. The authors describe meticulously the first part of the protocol until the sender part of the neuron performs Bell type measurement on microtubules *A* and *B*. However after that their exposition becomes ambiguous and misleading concluding that the teleportation of the state of microtubule *A* to microtubule

*C* is complete "with one caveat"[32]. Since there is a probabilistic nature to the described process, according to Mavromatos *et al.* (2002) microtubule *C* may receive the exact copy of the state of microtubule *A* or it may receive a state which is a unitary transformation away from the original, that is why there should be a "hardwired" condition that when microtubule *C* receives the correct microtubule state it does nothing further, yet if it receives one of the other three states, it performs the correct unitary transformation upon microtubule *C* in order to obtain the correct state from microtubule *A*. The real biophysical problem is to explain how this second part of the teleportation protocol could be implemented in the brain.

At present it seems hard to imagine a neurobiological process in which the neuron performs the necessary measurement in Bell basis upon its microtubules and encodes the result in neuronal firing pattern. Furthermore, it is also hard to imagine how neurons perform arbitrary unitary transformations on the underlying microtubules. Though at present one might be sympathetic with such a hypothesis if applied to the cerebral cortex, still there is the need of direct biophysical mechanism that ensures bidirectional interaction between the electric field of neurons and their microtubules. On one hand the neuron must be able to measure and report the state of its microtubules; on the other hand the neuron must be able to perform arbitrary unitary transformations upon its microtubules.

Briefly summarizing this section, we could say that although quantum teleportation protocols certainly do not operate within retina, it is tempting to consider the possibility of microtubule state quantum teleportation in the cerebral cortex. However in order to substantiate this hypothesis one must provide a feasible model for bi-directional interaction between neuronal microtubules and the neuronal electric field. This is needed for successful completion of both stages of the quantum teleportation protocol described in detail in *Appendix I. Until that time comes, we have*

---

[30] Note that the layers 2, 3 and 5 are clearly segregated from layers 1, 4, 6. Convergence of the segregated pathways requires projection to the same LGN layer (lamina) of both ipsilateral and contralateral pathways.

[31] Actually one needs transfer of mass or energy in order to communicate at least 2 bits of information. For details see *Appendix*. What Mavromatos and co-workers mean is that the system is not *transported* in space and time itself. Rather it is destroyed at one place in order to be re-assembled again at another place.

[32] The *Appendix* of this article fixes this ambiguity and provides detailed description of the second part of the protocol requiring the communication of at least 2 bits of classical information.

*to admit that we remain in the realm of fantasy, and that currently available quantum models of mind have nothing to do with the functioning of the real brain.*

## 4. Discussion

The input of visual information to the brain cortex is a multistage process, in which the initial stimulus is registered by photoreceptor cells, converted into electric currents that affect the membrane potential and subsequently into altered release of neuromediator (glutamate) through exocytosis. The bipolar, horizontal and amacrine cells process the obtained information using graded potentials, while the ganglion cells and the neurons from LGN process the converted into action potentials information using a kind of Boolean binary logic. Entering the brain cortex (V1) visual information is not equivalent pixel by pixel to the visual image entering the retina, and this is proved by some visual illusions (*Figure 3*). This shows that visual photons cannot collapse in the brain cortex; instead they must have collapsed in the retina.

Several researchers (Hameroff, 1998; Hilaire *et al.*, 2002; Gao, 2003; Salari *et al.*, 2008; Rahnama *et al.*, 2009) suppose that the informational input to the sensory organs is directly converted into quantum coherent states, thus missing the biological importance of pre-cortical processing and create hypotheses that conflict the clinical data. The experiments with implanting electrodes directly into the brain cortex (*Section 2.5*) suggest that the brain cortex is the residence for conscious experience. This notion is also well supported by clinical data. For example, lesions in the primary visual cortex cause cortical blindness (*amaurosis corticalis*), however in this case the pupillar reactions are preserved i.e. although there is a subcortical neural processing of information up to thalamus, it is not consciously realized because it does not enter the cerebral cortex and the patient insists that he/she cannot see anything. It can also be concluded that relevant stimulus for the cortical neurons is the electric current associated with changes in the neuronal membrane potential. Whether the cortex itself can sustain quantum coherent states relevant to consciousness (such as coherent microtubules) is an open question.

At the end, we would like to explain why the current work should be viewed as a significant step forward in developing new quantum mind theories consistent with biological data. With the birth of Karl Popper's critical rationalism it became clear that better scientific theories do not simply negate all the knowledge accumulated and explained by their predecessor theories. In contrast, they keep all (or most of) the previous information and then add some extra information. Therefore before one starts working on quantum mind, and quantum visual perception in particular, it is needed to have mastered to a certain (advanced) degree both neuroscience and quantum theory. Otherwise the hasty researcher is doomed to build a sand tower.

## Appendix I
### Quantum teleportation of qubits

The first *quantum teleportation* protocol, in which the quantum state of a qubit is transferred to another qubit with the use of a *maximally entangled*[33] qubit pair and a classical channel of information, was proposed by Bennett *et al.* (1993). The quantum teleportation does not transfer the quantum system itself in space-time. However transfer of energy indeed does occur and in the case of a two-level qubit teleportation one needs a classical channel of information capable to transmit at least two bits of information. Moreover, the quantum teleportation scheme neither allows for creation of multiple quantum copies of the teleported qubit, nor does allow for superluminal communication.

In order to please the curious reader, in the text following we will describe briefly the mechanism of quantum teleportation. First, we assume that a sender called *Alice* wants to teleport a simple two-level qubit *C* to a distant location where is positioned a receiver called *Bob*. In addition, we assume that Alice is provided with a classical channel of information capable to transmit messages to Bob, and that both Alice and Bob share respectively qubits *A* and *B*, each being a member of a maximally entangled qubit pair.

---

[33] *Maximally entangled* refers to one of four Bell states given below.

The *unknown quantum state* $|\psi\rangle_C$ of the qubit $C$ could be written as:

(1) $\qquad |\psi\rangle_C = \alpha|0\rangle_C + \beta|1\rangle_C$

Here the ket $|\psi\rangle_C$ is a column vector in a two-dimensional Hilbert space and can be alternatively written as:

(2) $\qquad |\psi\rangle_C = \begin{pmatrix} \alpha \\ \beta \end{pmatrix}$

In general Alice cannot measure for sure the coefficients $\alpha$ and $\beta$ of the unknown state due to the fact that *the quantum state is not observable* and thus not perfectly accessible for external observer. Nevertheless, this is irrelevant for the realization of effective quantum teleportation as we shall see below.

The quantum state of the maximally entangled qubit pair shared between Alice and Bob could be any of the following four Bell states:

(3) $\qquad |\Phi^+\rangle_{AB} = \frac{1}{\sqrt{2}}\big(|0\rangle_A \otimes |0\rangle_B + |1\rangle_A \otimes |1\rangle_B\big)$

(4) $\qquad |\Phi^-\rangle_{AB} = \frac{1}{\sqrt{2}}\big(|0\rangle_A \otimes |0\rangle_B - |1\rangle_A \otimes |1\rangle_B\big)$

(5) $\qquad |\Psi^+\rangle_{AB} = \frac{1}{\sqrt{2}}\big(|0\rangle_A \otimes |1\rangle_B + |1\rangle_A \otimes |0\rangle_B\big)$

(6) $\qquad |\Psi^-\rangle_{AB} = \frac{1}{\sqrt{2}}\big(|0\rangle_A \otimes |1\rangle_B - |1\rangle_A \otimes |0\rangle_B\big)$

In the following exposition we assume that Alice and Bob share entangled qubit pair, which is in the state $|\Phi^+\rangle_{AB}$. The composite three particle state is therefore:

(7) $\qquad \begin{aligned} &|\Phi^+\rangle_{AB} \otimes |\psi\rangle_C = \\ &= \frac{1}{\sqrt{2}}\big(|0\rangle_A \otimes |0\rangle_B + |1\rangle_A \otimes |1\rangle_B\big) \otimes \big(\alpha|0\rangle_C + \beta|1\rangle_C\big) \end{aligned}$

Alice is going to perform partial measurement of the two bits $A$ and $C$ in Bell basis:

(8) $\qquad |\Phi^+\rangle_{AC} = \frac{1}{\sqrt{2}}\big(|0\rangle_A \otimes |0\rangle_C + |1\rangle_A \otimes |1\rangle_C\big)$

(9) $\qquad |\Phi^-\rangle_{AC} = \frac{1}{\sqrt{2}}\big(|0\rangle_A \otimes |0\rangle_C - |1\rangle_A \otimes |1\rangle_C\big)$

(10) $\qquad |\Psi^+\rangle_{AC} = \frac{1}{\sqrt{2}}\big(|0\rangle_A \otimes |1\rangle_C + |1\rangle_A \otimes |0\rangle_C\big)$

(11) $\qquad |\Psi^-\rangle_{AC} = \frac{1}{\sqrt{2}}\big(|0\rangle_A \otimes |1\rangle_C - |1\rangle_A \otimes |0\rangle_C\big)$

In order to make clear the result of such a measurement we can re-write the three particle state using the following identities:

(12) $\qquad |0\rangle_A \otimes |0\rangle_C = \frac{1}{\sqrt{2}}\big(|\Phi^+\rangle_{AC} + |\Phi^-\rangle_{AC}\big)$

(13) $\qquad |1\rangle_A \otimes |1\rangle_C = \frac{1}{\sqrt{2}}\big(|\Phi^+\rangle_{AC} - |\Phi^-\rangle_{AC}\big)$

(14) $\qquad |0\rangle_A \otimes |1\rangle_C = \frac{1}{\sqrt{2}}\big(|\Psi^+\rangle_{AC} + |\Psi^-\rangle_{AC}\big)$

(15) $\qquad |1\rangle_A \otimes |0\rangle_C = \frac{1}{\sqrt{2}}\big(|\Psi^+\rangle_{AC} - |\Psi^-\rangle_{AC}\big)$

The three particle state is therefore:

(16) $\qquad \begin{aligned} &|\Phi^+\rangle_{AB} \otimes |\psi\rangle_C = \\ &\quad \frac{1}{\sqrt{2}}\alpha|0\rangle_A \otimes |0\rangle_B \otimes |0\rangle_C + \\ &\quad +\frac{1}{\sqrt{2}}\alpha|1\rangle_A \otimes |1\rangle_B \otimes |0\rangle_C + \\ &\quad +\frac{1}{\sqrt{2}}\beta|0\rangle_A \otimes |0\rangle_B \otimes |1\rangle_C + \\ &\quad +\frac{1}{\sqrt{2}}\beta|1\rangle_A \otimes |1\rangle_B \otimes |1\rangle_C \end{aligned}$

which after substitution becomes:

(17) $\qquad \begin{aligned} &|\Phi^+\rangle_{AB} \otimes |\psi\rangle_C = \\ &\quad \frac{1}{2}\big(|\Phi^+\rangle_{AC} + |\Phi^-\rangle_{AC}\big) \otimes \alpha|0\rangle_B + \\ &\quad +\frac{1}{2}\big(|\Psi^+\rangle_{AC} - |\Psi^-\rangle_{AC}\big) \otimes \alpha|1\rangle_B + \\ &\quad +\frac{1}{2}\big(|\Psi^+\rangle_{AC} + |\Psi^-\rangle_{AC}\big) \otimes \beta|0\rangle_B + \\ &\quad +\frac{1}{2}\big(|\Phi^+\rangle_{AC} - |\Phi^-\rangle_{AC}\big) \otimes \beta|1\rangle_B \end{aligned}$

and after re-arrangement of the terms simplifies to:

(18) $\qquad \begin{aligned} &|\Phi^+\rangle_{AB} \otimes |\psi\rangle_C = \\ &\quad \frac{1}{2}|\Phi^+\rangle_{AC} \otimes \big(\alpha|0\rangle_B + \beta|1\rangle_B\big) + \\ &\quad +\frac{1}{2}|\Phi^-\rangle_{AC} \otimes \big(\alpha|0\rangle_B - \beta|1\rangle_B\big) + \\ &\quad +\frac{1}{2}|\Psi^+\rangle_{AC} \otimes \big(\beta|0\rangle_B + \alpha|1\rangle_B\big) + \\ &\quad +\frac{1}{2}|\Psi^-\rangle_{AC} \otimes \big(\beta|0\rangle_B - \alpha|1\rangle_B\big) \end{aligned}$

After Alice measures the qubits $A$ and $C$ in Bell basis, the qubit $B$ evolves into mixture of four possibilities. In order for Bob to correctly restore the state $\alpha|0\rangle_B + \beta|1\rangle_B$ he needs to have the information of the measurement performed by Alice.

If Alice has obtained the state $|\Phi^+\rangle_{AC}$ then Bob should apply the identity operator $I \equiv \begin{bmatrix} 1 & 0 \\ 0 & 1 \end{bmatrix}$ to the qubit $B$ being in a state $\begin{pmatrix} \alpha \\ \beta \end{pmatrix}$, which actually means to do nothing since the state has been already teleported from qubit $C$ to qubit $B$.

If Alice has obtained the state $|\Phi^-\rangle_{AC}$ then Bob should apply the Pauli operator $\sigma_z \equiv \begin{bmatrix} 1 & 0 \\ 0 & -1 \end{bmatrix}$ to the qubit $B$ being in a state $\begin{pmatrix} \alpha \\ -\beta \end{pmatrix}$, which transforms the qubit $B$ into state $\begin{pmatrix} \alpha \\ \beta \end{pmatrix}$.

If Alice has obtained the state $|\Psi^+\rangle_{AC}$ then Bob should apply the Pauli operator $\sigma_x \equiv \begin{bmatrix} 0 & 1 \\ 1 & 0 \end{bmatrix}$ to the qubit $B$ being in a state $\begin{pmatrix} \beta \\ \alpha \end{pmatrix}$, which transforms the qubit $B$ into state $\begin{pmatrix} \alpha \\ \beta \end{pmatrix}$.

And lastly, if Alice has obtained the state $|\Psi^-\rangle_{AC}$ then Bob should apply the Pauli operator $\iota\sigma_y \equiv \begin{bmatrix} 0 & -1 \\ 1 & 0 \end{bmatrix}$ to the qubit $B$ being in a state $\begin{pmatrix} \beta \\ -\alpha \end{pmatrix}$ in order to get the desired state $\begin{pmatrix} \alpha \\ \beta \end{pmatrix}$.

Here one should pay attention to the fact that the Bob cannot obtain any useful information about the original state of the qubit $C$, by measuring his qubit $B$, which is in one of four states uniformly chosen at random. The probability for the qubit $B$ to be in each state $\begin{pmatrix} \alpha \\ \beta \end{pmatrix}$, $\begin{pmatrix} \alpha \\ -\beta \end{pmatrix}$, $\begin{pmatrix} \beta \\ \alpha \end{pmatrix}$ or $\begin{pmatrix} \beta \\ -\alpha \end{pmatrix}$, is exactly ¼. Therefore Bob needs to know the outcome of the measurement performed by Alice. Moreover, if an eavesdropper happens to learn the classical information communicated by Alice, he cannot produce a copy of qubit $C$. Alice does not know what was the original state of qubit $C$, so she cannot broadcast such an information. Otherwise one would be able to violate the quantum no-cloning theorem and make multiple perfect copies of unknown quantum states. Instead what Alice sends as a message is which one of the four operators Bob should apply to his qubit $B$ in order to transform it into the original state of qubit $C$. Nothing more, nothing less. The eavesdropper does not possess Bob's qubit $B$, so he cannot reproduce the original state of the qubit $C$. Finally, if Bob transforms his qubit $B$ into the state

$$\begin{pmatrix} \alpha \\ \beta \end{pmatrix},$$

no second copy of the qubit $C$ is achieved because the original qubit $C$ is destroyed in the process. *The last point to be made is that the coefficients* *$\alpha$ and $\beta$ were not available as a knowledge neither to Alice, nor to Bob or any other third party.* This is the essence of quantum teleportation and it cannot be used to send messages faster than light due to the requirement for a classical (relativistic) communication channel.

## Appendix II
## Mathematical sketch of Gao's argument

Here we outline briefly Gao's argument and try to translate the original *buzzword style*, into simple for understanding text. First let us have a two-level system (qubit) with two orthogonal states $\psi_1$ and $\psi_2$ [34]. We let the qubit states to be distinguished are the following non-orthogonal single states $\psi_1$ and $\frac{1}{\sqrt{2}}(\psi_1 + \psi_2)$.

The initial perception state of the observer is assumed to be $\chi$ and it is supposed that after interaction the corresponding entangled state of the whole system is respectively either $\psi_1\chi_1$ or $\frac{1}{\sqrt{2}}(\psi_1\chi_1 + \psi_2\chi_2)$, where $\chi_1$ and $\chi_2$ denote the perception state of the observer for the states $\psi_1$ and $\psi_2$. Gao assumes that the perception time of the observer for the definite state $\psi_1\chi_1$, which is denoted by $t_P$, is shorter than the dynamical collapse time for the superposition state $\frac{1}{\sqrt{2}}(\psi_1\chi_1 + \psi_2\chi_2)$, which is denoted by $t_C$, and the time difference $\Delta t = t_C - t_P$ is large enough for the observer to identify. In other words the observer is assumed to be able to subjectively feel, measure and tell the interval of time for which he arrived for example in state $\psi_1\chi_1$ (after the onset of the conscious measuring process). If this were true and the observer measured the state $\psi_1$, he could perceive his own state $\chi_1$ after time interval $t_P$ (plus he could feel and know that time $t_P$ has passed), while if the observer measured superposed state $\frac{1}{\sqrt{2}}(\psi_1 + \psi_2)$, he could perceive $\chi_1$ or $\chi_2$ only after the time interval $t_C$ (the observer had to wait for the superposed state to collapse to $\psi_1\chi_1$ or $\psi_2\chi_2$ plus he could feel and know that time $t_C$ has passed). In essence, since the observer is assumed to know (feel or measure) the time difference between $t_P$ and $t_C$, he could distinguish the measured nonorthogonal single states $\psi_1$ and $\psi_2$. Simply knowing that he arrived in state $\chi_1$ for shorter time $t_P$ and not for the longer time $t_C$ (starting from the onset of the measurement process), the observer could logically deduce e.g. that he came directly into this state and not going via intermediate superposed state, therefore the qubit state should have been a pure basis state $\psi_1$ and not the superposition of the two basis states $\frac{1}{\sqrt{2}}(\psi_1 + \psi_2)$.

---

[34] The wavefunctions $\psi_1$ and $\psi_2$ can be also written as vectors $|\psi_1\rangle$ and $|\psi_2\rangle$ in two-dimensional complex Hilbert space $\mathrm{H}$.

The validity of Gao's method relies on two conditions:

The first condition is that the perception time of the conscious being for the definite state e.g. $\psi_1$ is shorter than the perception time of the same state e.g. $\psi_1$ but following the dynamical collapse of the superposed state $\frac{1}{\sqrt{2}}(\psi_1 + \psi_2)$. To clarify the meaning of this condition let us write the perception process for the definite state $\psi_1$ as:

(P1)          $\psi_1 \rightarrow \psi_1 \chi_1$

Here $\psi_1 \chi_1$ expresses the fact that the conscious being has perceived $\psi_1$. The duration for this process (P1) is $t_P$. In contrast, for the perception of the same state $\psi_1$ but following the dynamical collapse of the superposed state $\frac{1}{\sqrt{2}}(\psi_1 + \psi_2)$, Gao has in mind the following process:

(P2)          $\frac{1}{\sqrt{2}}(\psi_1 + \psi_2) \rightarrow \frac{1}{\sqrt{2}}(\psi_1 \chi_1 + \psi_2 \chi_2) \rightarrow \psi_1 \chi_1$

This sequence of events implies that the conscious being evolves into an intermediate superposition state before arriving at the state $\psi_1 \chi_1$, therefore arriving at the state $\psi_1 \chi_1$ should take more time compared to process (P1). The duration for this process (P2) is $t_C$.

The second condition is that the time difference between the durations of the two processes described as (P1) and (P2), is long enough for the observer to identify (in other words the observer is aware of how much time has passed starting from the onset of the measurement process before perceiving e.g. $\psi_1$).

Gao (2003) proposes the following experiment to test his idea. He assumed that it is in principle possible to input a superposed state

$$\frac{1}{\sqrt{2}}(\psi_1 + \psi_2)$$

to a conscious being, which satisfies the above unusual conditions for time perception. The state

$$\frac{1}{\sqrt{2}}(\psi_1 + \psi_2)$$

is supposed to be a superposition state of a small number of photons that enter the eyes of the conscious being. After interaction the resulting entangled state of the whole system (photon plus human observer) is

$$\frac{1}{\sqrt{2}}(\psi_1 \chi_1 + \psi_2 \chi_2),$$

where $\chi_1$ and $\chi_2$ are the observer perception states for $\psi_1$ or $\psi_2$, respectively. If the conscious being satisfies the condition for time perception, he could distinguish the input states $\psi_1$ and $\frac{1}{\sqrt{2}}(\psi_1 + \psi_2)$ by knowing the time difference $\Delta t = t_C - t_P$.

Gao (2003) concludes that such distinguishing of non-orthogonal quantum states cannot be explained without referring to consciousness, so the physical world is not causally closed without consciousness. He also wrongly believes that experiments with photons could be helpful in proving the thesis above. In *Section 3.2.2* we have already provided detailed explanation why photons entering the eye and the retina of human observers cannot be used to deliver quantum coherent photon states to the brain cortex (and the human mind), instead what will be delivered as a visual information will be "collapsed" or "decohered" photon states. In addition, in a previous work we have already pointed out the fact that humans cannot discern time intervals shorter than 40 ms and we have discussed the implications of this fact for Q-mind theories (Georgiev, 2004).